# The validity of 21 cm spin temperature as a kinetic temperature indicator in atomic and molecular gas

Gargi. Shaw[1,2], G. J. Ferland[3], I. Hubeny[4]

## Abstract

The gas kinetic temperature ($T_K$) of various interstellar environments is often inferred from observations that can deduce level populations of atoms, ions, or molecules using spectral line observations; H I 21 cm is perhaps the most widely used with a long history. Usually the H I 21 cm line is assumed to be in thermal equilibrium and the populations are given by the Boltzmann distribution. A variety of processes, many involving Lyman alpha (Lyα), can affect the 21 cm line. Here we show how this is treated in the spectral simulation code Cloudy, and present numerical simulations of environments where this temperature indicator is used, with a detailed treatment of the physical processes that determine level populations within $H^0$. We discuss situations where this temperature indicator traces $T_K$, cases where they fail, as well as the effects of Lyα pumping on the 21 cm spin temperature. We also show that the Lyα excitation temperature rarely traces the gas kinetic temperature.

[1] Dept. of Physics, UM-DAE Centre for Excellence in Basic Sciences, University of Mumbai, Mumbai 400098, India; gargishaw@gmail.com

[2] Tata Institute of Fundamental Research, Homi Bhabha Road, Navy Nagar, Colaba, Mumbai 400005, India

[3] Department of Physics and Astronomy, University of Kentucky, Lexington, KY 40506; gary@uky.edu

[4] Steward Observatory, University of Arizona, Tucson, AZ 85721; hubeny@as.arizona.edu

*Subject headings:* radiative transfer, ISM: clouds, radio lines: galaxies

# 1 Introduction

The gas kinetic temperature, $T_K$, is one of the most important parameters for a comprehensive understanding of the interstellar and intergalactic medium (ISM & IGM: Spitzer 1978) and there are various ways to determine $T_K$ in neutral and molecular regions (Spitzer 1978; Tielens 2005; Osterbrock & Ferland 2006; Draine 2011). The 21 cm line originates from the transition between the hyperfine levels within the 1*S* level of $H^0$ (Griffith 1995). This line is a direct probe of conditions in the neutral atomic phase of the ISM (Heiles & Troland 2004, Murray et al. 2015).

The level populations are determined by thermal collisions, which drive the levels into equilibrium with the surrounding gas, and by non-thermal radiative processes, including direct pumping by the 21 cm radio continuum, pumping by the continuum near Lyman alpha (Lyα), and by the Lyα line itself (Field 1959; Urbaniak & Wolfe 1981; Deguchi & Watson 1985; Liszt 2001). It is well known that if the level populations are collisionally dominated in the absence of other non-thermal processes, the 21 cm line will trace $T_K$. In general, it is not possible to tell whether the 21 cm spin temperature (hereafter $T_s$), measured using emission and absorption in the case of neutral gas, traces $T_K$.

There are a few sources with sufficient observations to allow the comparison of the spin and kinetic temperatures from radio and UV observations. Orion's Veil, several layers of largely neutral gas lying between us and the main ionizing stars of the Orion Nebula, is an example where the spin temperature is much greater than the kinetic temperature (Abel et al. 2006; 2006). This also occurs in the damped Lyman alpha absorbers (DLAs) at $z = 1.78$ towards Q1331+ ($T_s \sim 1000$ K (Chengalur et al. 2000) and $T_K \sim 150$ K (Cui et al. 2005). These observations are difficult to make and hence there are relatively only a few situations where these comparisons between $T_s$ and $T_k$ can be made. This paper describes how we have incorporated the physics of the 21 cm transition into the spectral simulation code Cloudy, last described by Ferland et al (2013). Most processes that drive the 21 cm away from thermal equilibrium involve Lyα. Cloudy includes complete calculations of the ionization and level populations of hydrogen, including the effects of photoionization, cosmic rays, thermal excitation, and many other processes. The result is a self-consistent calculation of the Lyα intensity, appropriate for the local conditions. Many previous calculations assumed that the Lyα excitation temperature would be equal to the gas kinetic temperature if the Lyα line is optically thick (Wouthuysen 1952; Field 1959a, b; Chen & Miralda-Escude 2004). Actually the upper $2P$ level of $H^0$ is populated by non-thermal processes in the neutral ISM, especially cosmic ray excitation and recombination following cosmic ray ionization, so we find that the excitation temperature of Lyα is generally significantly greater than the local kinetic temperature. This drives the 21cm spin temperature away from the kinetic temperature and so it can only be treated with a fully self-consistent calculation.

This paper is organized as follows. The next section outlines our treatment of the micro-physics of the 21 cm line. We then present calculations of several standard environments to examine the 21 cm spin temperature and whether it traces the kinetic temperature. Further details can be found in the PhD thesis of Gargi Shaw, AAT 3198320. DAI-B 66/12, Jun 2006)

## 2 Calculations

This section describes the detailed microphysics included in our simulations. The calculation described below is a part of Cloudy since 2006 and has been briefly mentioned in Pellegrini et al. (2007), but has never been described in detail. All numerical calculations are performed by using version C13.04 of the numerical spectra simulation code Cloudy, which was last described by Ferland et al. (1998; 2013). Cloudy is based on a self-consistent calculation of the thermal, ionization, and chemical balance. Our chemical network, $H_2$ chemistry, and grain physics is described in Abel et al. (2005b), Shaw et al. (2005), and van Hoof et al. (2004) and summarized in Ferland et al. (2013). Our representation of H I line-formation processes is described by Ferland & Rees (1988), Ferguson & Ferland (1997), Luridiana et al. (2009), and Ferland et al. (2013). We considered both external and locally generated Ly$\alpha$ photons. Internal Ly$\alpha$ photons are created by recombination following ionization or by collisional deexcitation, by thermal or suprathermal particles, and may scatter many times before they either escape or are destroyed by dust (we treat this process following Netzer et al. 1985).

### 2.1 The 21 cm spin temperature ($T_s$)

The detailed mathematical procedure to compute $T_s$ is described below. Two notations for temperatures will be used, a local temperature designated by $t$ at each point of the cloud and an integrated temperature $T$ over that line of sight.

The ground state of $H^0$ undergoes hyperfine splitting due to the interaction between the magnetic moments of the proton and the electron spin creating $_1S_{1/2}$ (triplet) and $_0S_{1/2}$ (singlet) levels. Here $_FL_J$ notation is used to denote the hyperfine levels (Deguchi & Watson 1985) where F denotes the hyperfine quantum number. The statistical weight of a level with hyperfine quantum number $F$ is $(2\times F + 1)$ and the corresponding selection rule is $\Delta F = 0, \pm 1$. The levels $_1S_{1/2}$ and $_0S_{1/2}$ are separated by energy gap equivalent to 0.068 K. As summarized in Table 1, the hyperfine levels of 1S and 2P states, $_0S_{1/2}$, $_1S_{1/2}$, $_1P_{1/2}$, and $_1P_{3/2}$, are represented by levels 1, 2, 3 and 4 respectively (see also Figure 1 of Urbaniak & Wolfe 1981). Using this notation, the 21 cm transition is the 2-1 transition while Ly$\alpha$ has the transitions between 3 or 4 and 1 or 2 according the selection rule. The corresponding transition probabilities are calculated from equation 17.64 of Cowan (1981). Table 1 lists all these Einstein's $A$ coefficients. The following discussion focuses on the 21 cm transition, but in Cloudy these levels are part of a much larger model of $H^0$, as described in Ferland et al. (2013).

We define $t_s$ and $T_s$ in terms of the hyperfine level populations ($n_1$ and $n_2$) and column densities ($N_1$ and $N_2$) as

$$t_s = -\frac{0.068}{\ln(n_2/3n_1)} \text{ K, and } T_s = -\frac{0.068}{\ln(N_2/3N_1)} \text{ K.} \qquad (1)$$

We work in terms of the dimensionless photon occupation number within the Ly$\alpha$ transition, defined as $\eta = \frac{\bar{J}}{2h\nu^3/c^2}$ where $\bar{J}$ is the mean intensity and $2h\nu^3/c^2$ is the volume of phase space for photons. With this definition, the pump rate is given by $B_{ul} \bar{J} = A_{ul} \eta$. In the escape probability approach, $\bar{J}$ and the line source function ($S_l$) are related by $\bar{J} = S_l(1-\beta_{2P1S})$ where $\beta_{2P1S}$ is the escape probability and $S_l = \frac{2h\nu^3}{c^2} \times \frac{n(2P)/g_u}{n(1S)/g_l - n(2P)/g_u}$ (Elitzur 1992). Combining all these, the final expression for $\eta$ becomes

$$\eta = \frac{n(2P)g_l}{n(1S)g_u - n(2P)g_l} \times (1-\beta_{2P1S}) \approx \frac{n(2P)}{3n(1S)} \times (1-\beta_{2P1S}). \qquad (2)$$

The expressions for the rates that set the level populations in the upper (2) and lower (1) levels of the 21 cm line are,

$$R_{12} = C_{12} + P_{12} + 3 \times \frac{A_{32}}{A_{31}+A_{32}} \times A_{31} \times \eta_{13} + 3 \times \frac{A_{42}}{A_{42}+A_{41}} \times A_{41} \times \eta_{14}$$
$$R_{21} = A_{21} \times (1-\beta_{21}) + C_{21} + P_{21} + \frac{A_{31}}{A_{31}+A_{32}} \times A_{32} \times \eta_{23} + \frac{A_{41}}{A_{41}+A_{42}} \times A_{42} \times \eta_{24} \qquad (3).$$

$R_{12}$ represents rates for all processes that depopulates level 1 and eventually populates level 2. The reverse rate is denoted by $R_{21}$. Here $C$ and $P$ represent rates for collisions and external pumping in the 21 cm line respectively. The hyperfine-resolved $\eta$'s are all nearly equal since the hyperfine splitting is small compared with the energy of Lyα. We assume that they are equal in the following but do account for the small energy differences in our code. The values of $\eta_{13}, \eta_{14}, \eta_{23}$, and $\eta_{24}$ are listed in Table 1. The terms $3 \times \frac{A_{32}}{A_{31}+A_{32}} \times A_{31} \times \eta_{13}$ and $3 \times \frac{A_{42}}{A_{42}+A_{41}} \times A_{41} \times \eta_{14}$ represent the rate of level 2 being pumped from level 1 via levels 3 and 4 respectively. Similarly, $\frac{A_{31}}{A_{31}+A_{32}} \times A_{32} \times \eta_{23}$ and $\frac{A_{41}}{A_{41}+A_{42}} \times A_{42} \times \eta_{24}$ represent the rate of level 1 being pumped from level 2 via levels 3 and 4 respectively.

We include collisional excitation and de-excitation in the upper and lower level of 21 cm by various colliders such as $H^0$, $e^-$, and $H^+$. The collisional de-excitation rate coefficients for $H^0$ 21 cm are taken from Zygelman (2005) in the temperature range 1 K to 300K. We use Allison & Dalgarno (1969)'s rate multiplied by 1.24 above 300 K. Above 1000K, we extrapolate the rate as a power of $T^{0.33}$. The collisional rate coefficients for electrons are taken from Smith (1966) and updated to the values given by Liszt (2001), whereas the rate coefficient for $H^+$ is taken from Furlanetto et al. (2007).

If the 1$S$ populations are determined by processes that are in detailed balance, such as collisions, they will have a Boltzmann distribution and so determine the kinetic temperature. However, other processes, mainly pumping by the Ly$\alpha$ continuum, Ly$\alpha$, or the CMB, or suprathermal electron excitation following cosmic ray or x-ray ionization, can affect the populations, and cause $T_s \neq T_k$ (Field 1959)

The radiative transition rate between the two hyperfine levels is $2.85 \times 10^{-15}$ s$^{-1}$ (Wild 1952), which is usually small compared to the collision rates given above, $2.1 \times 10^{-10}$ $n(H^0)$ s$^{-1}$. The critical density, the density were collisions are more important than radiative decays, is $n(H^0) = 1.37 \times 10^{-5}$ cm$^{-3}$ for $T = 300$ K. This is very low and hence the level populations will trace $T_K$ if they were determined solely by collisional processes and radiative decays.

The populations of hyperfine levels can be affected by both line and continuum pumping, which can drive $t_s$ away from $t_K$. Pumping by the external continuum, usually the CMB, is treated as in Ferland (1991). This is very important before the re-ionization epoch when stars have not yet formed. Pumping by the external Ly$\alpha$ continuum is also very important and the pumping rate [s$^{-1}$] is simply $R_{lu} = J_{lu} B_{lu} = \eta_{ly\alpha} A_{ul} g_u/g_l$ (s$^{-1}$). The rate of pumping by the external continuum depends on the line optical depth and resulting self-shielding and finally can be expressed as

$$\eta = \eta_{CMB} \times \beta_{21cm}.$$

## 2.2. *The effects of Lyα*

The following presents simple estimates of the effects of Lyα upon 21 cm. The intensity of the Lyα line *J* is set by the 1*S* and 2*P* level populations. We derive simple estimates of these populations and the Lyα occupation number by assuming that n = 2 is mainly populated following recombination, as shown by the detailed numerical simulations presented below. For simplicity the following equations neglect terms accounting for direct collisional excitation from 1*S* to 2*P* since these are negligible for the low temperatures present in the neutral ISM. In the numerical simulations the populations are determined from the simultaneous solution of the hydrogen ionization and level populations as discussed in Ferland et al. (2013).

We present the Lyα excitation temperature by $t_{Lya}$ as

$$t_{Lya} = -\frac{1.183 \times 10^5}{\ln\left[n(2P)/3n(1S)\right]}. \tag{4}$$

Although we actually solve a full system of equations, which includes a large number of physical processes, we can simplify our results in the low-density ISM limit to illustrate the important effects. The hydrogen ionization equilibrium can be written as

$$n(1S)\Gamma = n_e n_p \alpha_B, \tag{5}$$

(Osterbrock & Ferland 2006) and the level populations of *n*(1*S*) and *n*(2*P*) are derived from

$$n(2P)[C_{2P1S} + A_{2P1S}(\beta_{2P1S} + \delta)] = n_e n_p \alpha_B(2P). \tag{6}$$

Here $\Gamma$, $n_e$, $n_p$, $\alpha_B$, $C_{2P1S}$, $A_{2P1S}$, and $\delta$ represent the ionization rate (both photo and cosmic ray ionization), the electron and proton densities, the Case B radiative recombination rate coefficient, the collisional de-excitation rate, the transition probability, and the destruction probability, respectively. In an H II region photoionization is the dominant ionization process while cosmic-ray and x-ray ionization are important in the neutral and molecular ISM. In both cases Lyα is produced by recombination. For the cosmic ray ionization, we assume an $H^0$ cosmic ray ionization rate of $2 \times 10^{-16}$ s$^{-1}$ as the Galactic background value in atomic regions (McCall et al. 2003).

Equations (5) and (6) give the ratio of level populations, which then determines $\eta_{ly\alpha}$, and thus the pumping rate between the hyperfine levels of the 1S state. Hence,

$$\eta_{lya} \approx \frac{n(2P)}{3n(1S)} \times (1-\beta_{2P1S}) \approx \frac{\alpha_{eff}(2P)}{3\alpha_\beta} \times \frac{\Gamma}{C_{2P1S} + A_{2P1S}(\beta_{2P1S}+\delta)} \times (1-\beta_{2P1S}) \qquad (7)$$

where we assume $n(2P) << n(1S)$. In the ISM limit for a UV transition like Lyα, where $h\nu > kT_k$, $t_{lya}$ and $\eta_{lya}$ are simply different forms of the population ratio.

## 2.3. Simple estimates

The following section will present numerical results for the 21 cm spin temperature and the Lyα excitation temperature. The slope of the Lyα source function, and not the intensity, sets the 21 cm spin temperature when pumping dominates collisions and cosmic rays. In the calculations presented in this paper, we assume constant source function at Lyα line center (For detail see Appendix).

The 2P term is populated following radiative recombination and depopulated by radiative and collisional decays to 1S. Once created, Lyα photons undergo a large number of scatterings and either escape the cloud, are absorbed by dust, or are collisionally deexcited (thermalized). The scattering and absorption rates are the fastest in ISM conditions because the critical density of 2P at 70K, $n_{crit} = [A_{2P1S}(\beta_{2P1S}+\delta)]/C_{2P1S} \approx 10^{15}(\beta_{2P1S}+\delta)$ [cm$^{-3}$] is large. The line will be collisionally deexcited only when $n > (\beta_{2P1S} + \delta)10^{15}$. Here $\delta$ depends on the ratio of the absorption to scattering plus absorption probabilities of dust and Ly$\alpha$ through (Netzer et al. 1985),

$$\delta = \frac{n_d \sigma_d}{n_d \sigma_d + n_{H^0} \sigma_{Ly\alpha}}. \tag{8}$$

Here $n_d$, $\sigma_d$, $n_H^0$, $\sigma_{Ly\alpha}$ represent the dust density, dust absorption coefficient, neutral hydrogen density, and Ly$\alpha$ absorption coefficient, respectively. As shown below, $\delta$ dominates in atomic regions of the ISM, so most Ly$\alpha$ photons are lost by dust absorption rather than by escape or collisional deexcitation. Although the numerical calculations will include all terms, the following analytical estimates will consider only the escape and absorption terms.

The ISM near a newly formed star can be in three different gas phases, an H II region with H$^+$, a PDR (photo-dissociation region) with H$^0$ and H$_2$, and a molecular region with H$_2$. We next describe the characteristics of $\eta_{Ly\alpha}$, or equivalently the 2P/1S population ratio [see equation 7], for each of these phases.

We first consider the H II region. The Ly$\alpha$ optical depth is small near the illuminated surface of a cloud, so $\beta_{2P1S} \sim 1$ and line photons escape freely. As a result, $\eta$ is small and $t_{Ly\alpha}$ is large. Deep in the H II region the Ly$\alpha$ line becomes optically thick and is strongly trapped, $\beta_{2P1S}$ decreases while $\Gamma$ is also large. As a result, $\eta_{Ly\alpha}$ becomes larger and $t_{Ly\alpha}$ becomes smaller. Furthermore, since little H$^0$ is present, the line opacity is small, the absorption to scattering ratio is large, and the destruction probability is large. In this case $\eta_{Ly\alpha}$ simplifies to

$$\begin{aligned}\eta_{Ly\alpha} &= \frac{\alpha_{eff}(2P)}{3\alpha_\beta} \times \frac{\Gamma}{A_{2P1S}\delta} \\ &= \frac{\alpha_{eff}(2P)}{3\alpha_\beta} \times \frac{\Gamma}{A_{2P1S}} \times \frac{n_{H^0}\sigma_{Ly\alpha} + n_d\sigma_d}{n_d\sigma_d} \approx \frac{\alpha_{eff}(2P)}{3\alpha_\beta} \times \frac{\Gamma}{A_{2P1S}}\end{aligned}. \tag{9}$$

Hydrogen is atomic and molecular in the PDR. $\Gamma$ decreases in the transition from the ionized region, where it is dominated by photoionization, to the atomic and molecular regions, where cosmic ray or x-ray ionization dominates. The $H^0$ fraction is large in the PDR, so $n_{H^0}\sigma_{Ly\alpha} \gg n_d\sigma_d$, and therefore, $\eta_{Ly\alpha} = \frac{\alpha_{eff}(2P)}{3\alpha_\beta} \times \frac{\Gamma}{A_{2P1S}} \times \frac{n_{H^0}\sigma_{Ly\alpha}}{n_d\sigma_d}$. The smaller $\Gamma$ in the PDR causes $\eta$ to be smaller and Ly$\alpha$ pumping becomes less important and $t_{Ly\alpha}$ decreases further. Finally, in molecular regions, where hydrogen is mostly $H_2$ and little $H^0$ is present, $n_{H^0}\sigma_{Ly\alpha} \ll n_d\sigma_d$. In this case, $\beta_{2P1S}$ is small, and consequently $\eta_{Ly\alpha} = \frac{\alpha_{eff}(2P)}{3\alpha_\beta} \times \frac{\Gamma}{A_{2P1S}}$. This clearly shows that $\eta_{Ly\alpha}$ is very small and pumping is insignificant in molecular gas when dust is present.

However, in a metal-free and dust-free environment the destruction probability $\delta$ due to absorption of Ly$\alpha$ photons by grains is zero. Then $\eta_{ly\alpha}$ is proportional to,

$$\eta_{Ly\alpha} \approx \frac{\alpha_{eff}(2P)}{3\alpha_\beta} \times \frac{\Gamma}{C_{2P1S} + A_{2P1S}\beta_{2P1S}}. \tag{10}$$

As a result, Ly$\alpha$ scattering and pumping is more important in dust-free environments.

## 3 Results – Various astrophysical environments

In this section we present simulations of M17 and Orion's Veil where observations have revealed differences between $T_s$ and $T_K$. We also present simulations for Galactic ISM. It concludes with a discussion of the effects of metallicity on the spin temperature in DLAs) like systems.

## 3.1 Orion's Veil

Orion's Veil is an absorbing screen lying along the line of sight of Orion H II region (O'Dell, 2001). This is the nearest H II region and is a region of active star formation, although the star formation rate is relatively low. The H II region is mainly ionized by a single O star. It is one of the few regions where a detailed map of the magnetic field strength exists from 21 cm circular polarization studies (Troland et al. 1989). There are two main components (Component A and B) lying within few parsecs of the Trapezium cluster with different turbulent and magnetic pressures. Abel et al. (2006) studied these layers extensively and derived density, kinetic temperature, and distance from the Trapezium, from analysis of various ionic, atomic and molecular lines together with 21 cm measurements of the line of sight component of the magnetic field. The density and magnetic field strength are both higher than that is typically found in the cold neutral medium (CNM), and the resulting gas pressure is also higher.

Here we have adopted the parameters derived by Abel et al. (2006) and recomputed their Veil model. The hydrogen density $n_H$ is ~$10^{2.5}$ cm$^{-3}$ for component A and ~$10^{3.4}$ cm$^{-3}$ for component B. They derived an upper limit on $T_k$ by assuming that the 21 cm H I line widths are entirely due to thermal motions and also used the density-temperature relationship from Abel et al. (2005) to deduce the same temperature ($T_K$ = 50 K and 80 K in component A and B of Orion veil). The 21 cm spin temperature was determined from the observed $N(H^0)/Ts$ with values $10^{19}$ cm$^{-2}$ K$^{-1}$ and $10^{19}$ cm$^{-2}$ K$^{-1}$ for component A and B, respectively. Hence, they concluded that for component A and component B, 80 K < $T_s$ < 110 K and 100 K < $T_s$ < 165 K, respectively.

Figure 1 shows the hydrogen ionization structure of component B of the Orion Veil. Hydrogen is mostly H$^+$ in the H II region; this is followed by an atomic hydrogen region (the PDR). In the H II region $t_K$ ~ 8000 K and the electron density is nearly equal to the H$^+$ density. In the PDR region $t_K$ is lower (~115 K) and $C^+$ is the main electron donor. The molecular fraction is small in the Veil due to its proximity to the Trapezium. The small separation and resulting high radiation field is the reason for the puzzlingly small molecular hydrogen abundance first noted by Savage et al. (1977).

Figure 2 shows the various temperature indicators as a function of depth into the Veil from its illuminated face. The ionization rate is high in the H II region and consequently both $\eta_{Ly\alpha}$ and $t_{Ly\alpha}$ are high (equation 7, section 2.2). As a result, $t_s$ is close to $t_{Ly\alpha}$ although less than $t_K$.

The kinetic temperature decreases sharply across the ionization front and reaches ~115K in the atomic region. The ionization rate decreases and the neutral hydrogen density increases across the PDR. As a result, Ly$\alpha$ pumping is less than that in the ionized region (section 2.2) and $t_s$ is less than <u>that</u> in the H II region. At the shielded face the spin temperature approaches the kinetic temperature and the difference between $t_k$ and $t_s$ is only 13 K. However $t_{Ly\alpha}$ remains much higher than $t_k$ due to the significant recombination contribution to the line.

The populations of the $H^0$ 2P term are affected by Ly$\alpha$ pumping by the external continuum at shallow depths into the $H^+$ layer, increasing $\eta_{ly\alpha}$. This, in turn affects the 21cm spin temperature. The external continuum near the Ly$\alpha$ line is set by the SED of the ionizing star. We use a modified Kurucz LTE stellar ionizing continuum atmosphere as described in Rubin et al. (2001) as the incident SED. In real stars the Ly$\alpha$ line could be in absorption or in emission, depending on the wind structure. The Ly$\alpha$ line is very difficult to observe directly since, in most cases, the line is absorbed by the intervening ISM. The detailed effect of the Ly$\alpha$ line upon 21 cm is determined by the stellar SED. The thick-black lines in Figure 3 show the various temperatures in the absence of continuum Ly$\alpha$ pumping. In this case, where we have assumed that the stellar SED has no emission in the Ly$\alpha$ line, $t_s$ and $t_k$ traces each other in the atomic region. The observed value of $T_s$ is higher than the $T_k$ deduced by Abel et al. (2006) implying that the star does increase the Lyα pumping. A similar effect was observed by Pellegrini et al. (2007) which was shown in their Appendix.

## 3.2 Orion H II region

The appendix of Pellegrini et al. (2007) shows the effects of pumping by Lyman lines in the stellar continuum upon the 21 cm spin temperature in the PDR. For the parameters of their model of M17 even a modest amount of radiation in the stellar Lyman lines could affect the 21 cm temperature by the processes described above.

As a further test, we have recomputed the Baldwin et al. (1991) model of the Orion H II region, extending it into the PDR and stopping at an extinction of $A_V = 10$. We follow the procedure described by Pellegrini et al (2007) in blocking the stellar Lyman lines, with the exception that we now allow the lines to go into emission (blocking factors > 1). The results are shown in Figure 4. The parameters for the Orion H II region are quite different from M17 and the results shown in the Figure differ too. The effects are modest for low blocking and although they become extreme if the Lyman lines are in emission. These results, and those of Pellegrini, show that a wide range of effects is possible.

## 3.3 M17

The Galactic H II region M17 is a relatively nearby massive region of star formation situated at a distance of $1.6 \pm 0.3$ kpc (Nielbock et al. 2001). An overview of the geometry and the physical conditions in the cloud is given in Pellegrini et al. (2007). Here we use the same geometry and physical conditions. In contrast with the Orion region, the H II region is created by a cluster of stars with a significant number of massive O stars. The ionizing radiation comes from a nearby dense cluster of O and B stars and an X-ray continuum. The southwestern (SW) part of M17 contains an obscured ionization front that is viewed nearly edge-on. It offers an excellent opportunity to study the way gas changes from fully ionized to molecular as radiation from the ionizing stars penetrates into the gas. It is a classic example of a blister H II region where the ionization front is moving into the nearby molecular region (Pellegrini et al. 2007).

Pellegrini et al. (2007) derived a self-consistent model using various ionic, atomic and molecular lines and accounted for the magnetic field in a magnetohydrostatic model. Their model is time-steady. The force caused by the outward momentum in starlight is balanced by pressures present in the nebula, including gas, magnetic, and turbulent pressure. The [S II] $\lambda$6717, 6731 doublet ratio was used to measure the density in the H II region. This star-forming region has a high magnetic field strength deduced from 21 cm circular polarization studies (Brogan et al. 1999; Brogan & Troland 2001) and the magnetic pressure dominates over the gas pressure. Pellegrini et al. (2007) also included a high cosmic-ray flux in their model as it was necessary to predict the observed line ratios of I([O I] $\lambda$ 146 µm)/I(H$\beta$) and the radio map of 330 MHz synchrotron emission. An extinction value of $A_V$ = 6.36 which is the observed extinction internal to M17, $A_V$ = 9 mag, multiplied by cos (45°) is used as the stopping criterion. $T_s$ was determined from the ratio of the neutral hydrogen column density $N(H^0)$ to the spin temperature = $9 \times 10^{19}$ cm$^{-2}$ K$^{-1}$, which was measured from observed optical depth and the emission brightness temperature.

Pellegrini et al. (2007) had used the theory presented here to derive spin temperatures but did not go into the detailed physics. Figure 5 shows the computed H-ionization structure. Hydrogen is mostly H$^+$ in the H II region; this is followed by a neutral region (the PDR). The electron density is highest in the H II region due to photoionization of hydrogen. The density is nearly equal to the density of $H^+$. The electron density decreases in the neutral region.$H^0$.

Figure 6 plots the various temperature-indicators as a function of depth from the illuminated face of the cloud. The kinetic temperature is $1.5\times10^4$ K at the illuminated face due to photoionization by the very hot O stars. The ionization rate is high in the H II region and consequently both $\eta_{Ly\alpha}$ and $t_{Ly\alpha}$ are high (equation 7, section 2.2). As a result, $t_s$ is close to $t_{Ly\alpha}$ although less than $t_K$. However, $t_K$ decreases sharply across the ionization front. The ionization rate decreases and the neutral hydrogen density increases across the PDR. As a result, Lyα pumping is less than that in the ionized region (section 2.2) and $t_s$ is less than that in the H II region. But still its value is higher than $t_k$. In this region, thus, $t_s$ does not trace $t_k$. In these ways M17 is much like a scaled up version of the Orion Veil discussed above.

## 3.4 Galactic ISM

This section discusses the kinetic and 21 cm spin temperatures for the Galactic ISM, which consists of the cold neutral medium (CNM) and warm neutral medium (WNM). This is an important test since many observational results are available (Heiles & Troland 2003; Murray et al. 2015 and references cited there). A particular line of sight might contain both CNM and WNM. But here we do not consider such combination. To model this environment, we made a grid of constant-density models with ISM abundances, graphitic and astronomical silicate grains, and the background ISM radiation field described by Black (1985). We vary the hydrogen volume density and total neutral hydrogen column density from 0.01 to 158 $cm^{-3}$ and $10^{19}$ to $10^{22}$ $cm^{-2}$ respectively.

Results are presented in Table 2 and Figure 7. Tables 2a, 2b, 2c and 2d list our predicted average kinetic temperature and average 21 cm spin temperature. These temperatures vary over a wide range. The average spin temperature varies from 3350 K to 40 K. Figure 7 shows the gas pressure as a function of hydrogen density for various total neutral hydrogen column densities. It shows that in addition to CNM and WNM, thermally unstable regions also exists in the Galactic ISM. Our simulations show that the average 21 cm spin temperature is lower than the average kinetic temperature for the WNM as collisions are much less in this low-density region. In the high-density low-temperature CNM, the average kinetic temperature is less than or equal to the averaged 21 cm spin temperature. In addition to these, we have also shown the 21cm optical depths in table 2. Our calculations show that the optical depth of 21cm is less than one for the WNMs of the clouds with $N(H\ I) < 10^{22}$ cm$^{-2}$.

## 3.5 Effect of metallicity on 21cm spin temperature

This section examines the effects of metallicity and the background cosmic ray ionization rate on the 21 cm spin temperature for clouds similar to damped Lyα absorbers. We find that the spin and kinetic temperatures are only similar when the cosmic ray ionization rate is low or the gas is dusty. These parameters have significant effects on the spin temperature.

To model a DLA with in-situ star-formation, we begin with parameters similar to the cold neutral medium (CNM) of the ISM and later decrease the metallicity and dust abundances to mimic such DLAs at high redshift (Srianand et al. 2005). The hydrogen density in CNM is ≥ 10 cm$^{-3}$ (Draine, B. T. 2011). In the previous subsection (3.4) we have shown that temperature depends on hydrogen density. As a simplest model, we considered plane parallel geometry for the gas cloud with hydrogen density 10 and 100 cm$^{-3}$. This gas cloud is irradiated by our standard ISM radiation field from one side and bathed in a z = 0 CMB. Earlier Black & van Dishoeck (1987) had used this type of one-sided radiation field in their PDR models. Furthermore, we have assumed a cosmic ray ionization rate of H$^0$ (CR) of $10^{-17.3}$ s$^{-1}$. This value is smaller than the current Galactic background CR rate (McCall et al. 2003). All the models stop at a neutral hydrogen column density $10^{21}$ cm$^{-2}$.

Figures 8a and 8b show the difference between $t_s$ (solid line) and the corresponding $t_K$ (dashed-line) for varying metallicity and dust content in the range of standard ISM to 0.01 ISM. Our simulation shows that at the shielded face of the cloud, $t_K$ and $t_s$ are nearly equal. For a cloud with density 100 cm$^{-3}$, these two temperatures are less than that of a cloud with density 10 cm$^{-3}$. As the metallicity and dust content decreases, cooling by metal lines decreases and hence both temperatures increase suggesting an anti-correlation between metallicity and $t_s$. This simulation has implications for high redshift intervening clouds since it is observed that both metallicity and dust abundances decrease as a function of redshift. We like to point out that recently Kanekar et al. (2014) measured high spin temperatures (≥ 1000 K) for a sample of 37 DLAs at high redshift and found that $T_s$ is anti-correlated with metallicity.

For this same model, in Figures 9a and 9b, we present a very important consequence of CR rate on the difference between $t_s$ and $t_K$. The solid and the dashed-line represent $t_s$ and $t_K$ like the previous figure. We varied the CR rate from $10^{-17.3}$ s$^{-1}$ to $10^{-15.3}$ s$^{-1}$. The difference between these two temperatures increases as a function of the CR rate. However, the difference between these two temperatures <u>is</u> smaller for higher density. High-energy cosmic rays produce secondary electrons which heat, ionize and excite H$^0$. Shull et al. (1985) show these rates for a wide range of ionization fraction, and that the ionization and excitation of H$^0$ is comparable. The Lyα production increases with CR rate and hence the bigger difference between $t_K$ and $t_s$. It might be erroneous to assume that $t_s$ traces $t_K$ in environments where the CR rate is not known accurately. However, this fact can be used as an advantage to constrain the CR rate and star formation rate.

Furthermore, the 21 cm line optical depth is inversely proportional to $T_s$, $\int \tau_{21} dv \propto N_{H\,I} \times [f/T_s]$, where f is the covering factor. For a given f, a high $T_s$ suggests a smaller $\tau_{21}$. As a consequence, it would be hard to detect 21 cm line in absorption in environments with low density when the CR rate is very high or the metallicity is very low.

## 4 Conclusions

The aim of the current work is to present our treatment of the micro-physics of 21 cm line in Cloudy, and show whether $t_s$ and $t_{lya}$ <u>trace</u> $t_k$ in various astrophysical environments. A major outcome of this work is that the common assumption that the Lyα excitation temperature ($t_{Ly\alpha}$) will trace $t_k$ if the gas is optically thick and the Lyα radiation field will drive $T_s$ to $T_k$. (Wouthuysen 1952;, Field 1959; and most recently Chen & Miralda-Escude 2004) is not valid for all astrophysical environments. $t_{Ly\alpha}$ depends on the ratio of populations of the 2*P* to 1*S* terms of $H^0$ and this ratio is mainly determined by the balance between creation of new Lyα photons following radiative recombinations, Lyα pumping, and excitation by suprathermal electrons, and their loss by collisional de-excitation and absorption by dust. These processes are not related by detailed balance, so will not drive level populations to local thermodynamic equilibrium. Hence $t_{Ly\alpha}$ is often not equal to $t_K$.

We arrive at the following conclusions:

- $t_{lya}$ does not trace $t_K$ since the $2P$ population is set by the balance between radiative recombination and downward collisions which are not related to one another by detailed balance (Figures 2, 3 and 6).

- In metal-poor dust-poor environments, $t_{lya}$ forces $t_s$ away from $t_k$ (Figures 8a, 8b).

- None of the temperature indicators traces the kinetic temperature in the ionized and atomic regions in M17 because of Lyα pumping (Figure 6).

- The effect of Lyα continuum pumping on 21cm spin temperature is important. In real stars the Lyα line can be in absorption or in emission. It is possible to predict the nature of Lyα continuum in stellar atmospheres if we have observed values of $T_s$ and $T_k$ (Figure 4).

- The difference between $t_s$ and $t_k$ increases as a function of CR which can be used in advantage of constraining CR rate (Figures 9a, 9b).

## 5 Acknowledgements

We thank Keith MacAdam, Phillip Stancil and R. Srianand for useful discussions. GJF acknowledges support by NSF (1108928, 1109061, and 1412155), NASA (10-ATP10-0053, 10-ADAP10-0073, NNX12AH73G, and ATP13-0153), and STScI (HST-AR- 13245, GO-12560, HST-GO-12309, GO-13310.002-A, and HST-AR-13914). We thank our anonymous referee for his/her thoughtful suggestions.

# 6 Appendix- The Wouthuysen-Field Effect

If Lyα pumping dominates the 21 cm level populations then the ratio of the populations of the upper and lower state of the 21 cm line, and the resulting spin temperature, is determined by the slope of the Lyα source function at line center. The literature on this is quite large. Among many others, Deguchi & Watson (1985, hereafter DW85) discuss this for the cold ISM and Hirata (2006) in the cosmological context. The slope of the Lyα source function, and not the intensity, sets the 21 cm spin temperature when pumping dominates. In the calculations presented in this paper, Lyα pumping was often the most important process affecting the 21 cm populations.

Three temperatures apply to H I Lyα in the ISM, a highly non-equilibrium environment. The *gas kinetic temperature* describes the thermal motions of particles in the gas. The Lyα *excitation temperature* is determined by the 2*p* / 1*s* level population ratio and the Boltzmann equation. The Lyα source function can be described in terms of these level populations, and is $S_v = \left(2h v^3 / c^2\right) \times n_u / \left(n_l g_u / g_l - n_u\right)$. Here *l* and *u* represent the lower and upper levels, and the *g*'s are statistical weights. From this definition, and the relationships between the Einstein A and B coefficients, it follows that the source function near line center is given by the Planck function at the excitation temperature. Finally, the Lyα *color temperature* can be defined as the temperature of the Planck function that describes the slope of the Lyα source function at line center. It is this color temperature that, through photoexcitation, strongly affects the 21 cm spin temperature when pumping dominates.

The Lyα pumping of 21 cm is frequently referred to as the Wouthuysen-Field effect, in recognition of the first two authors on the topic. The original Wouthuysen (1952) paper is the printed summary of his oral contribution to the 1951 meeting of the American Astronomical

Society in Cleveland OH. Wouthuysen notes that, in the limit of an infinite number of scatterings, the Lyα color temperature will go over to the kinetic temperature due to small changes in the photon energy produced by recoil of the atom as the photon is absorbed.

This infinite-scattering limit cannot occur at finite densities because the 2$P$ level will eventually be collisionally deexcited. In the limit of a very large number of scatterings collisional excitation and deexcitation will thermalize the 2$P$/1$S$ level populations and the Lyα excitation temperature will equal the gas kinetic temperature. So, in this limit, the Lyα color, excitation, and kinetic temperatures are all equal. Field's two seminal papers (Field 1959a, 1959b), which consider the large-scattering case, confirms that the color and kinetic temperatures are equal.

The ISM is far from equilibrium. The Lyα line is quite optically thick so line photons are mainly lost through absorption onto dust. The gas kinetic temperature for CNM is low, usually well below 1000 K. As a result, thermal collisional excitation and photo-pumping of Lyα are negligible, a major difference from the cosmological case. Instead, cosmic rays produce Lyα, both by direct excitation by suprathermal secondary electrons (Spitzer & Tomasko 1968) and by collisional ionization followed by recombination. The Lyα excitation temperature is generally an order of magnitude higher than the kinetic temperature, as Figure A1 shows.

We consider three cases for the color temperature. The definition of the line source function, together with relationships between the Einstein $A$ and $B$ coefficients, make the line source function equal to the Planck function at the Lyα excitation temperature. In this thermodynamic limit the color temperature will be equal to the excitation temperature. This is preferred by Elitzur & Ferland (1986), quoting Mihalas (1978). We refer to this as the *excitation* case.

Recoil might eventually bring the color temperature and kinetic temperature together, if enough scatterings occur, as argued by Wouthuysen & Field. We refer to this as the *kinetic* case.

Discussions of radiative transfer in the stellar atmospheres (Rutten 2000, Hubeny & Mihalas 2014) do not mention recoil as a significant process affecting line transfer or the line source function. Rather, in the case of complete redistribution, which applies in the core of the line, the source function is $S_v=$ *constant*. Hubeny & Mihalas (2014) review the general picture in Section 10.1 and give an example in Section 15.6. Rutten (2003) gives this result in the discussion around equation 2.73. The Lyα slope is non-thermal in the $S_v=$ *constant* case so cannot be described by a color temperature. We refer to this as the *constant* case.

Which case applies? Very few calculations of the Lyα source function include recoil since the effect is so small. Field (1959b) studied time relaxation of a resonance line profile without transfer and confirmed Wouthuysen's statement in the limit of many scatterings. Adams (1971) was the first study that quantified how many scatterings are needed for recoil to affect Lyα. Line photons receive a frequency shift of one Doppler width in each scattering. The recoil shift is about $10^4$ times smaller than the Doppler shift, so a very large number of scatterings <u>is</u> needed for the recoil energy shift to be significant. He concluded that recoil can be neglected when Lyα undergoes less than $5.6 \times 10^{10}$ scatterings. His limit is never reached in our calculations, where dust absorbs Lyα after $\sim 10^6$ - $10^7$ scatterings.

The later study by DW85 found that a much smaller optical depth of $10^5 - 10^6$, corresponding to roughly $10^6$ scatterings, will bring the color temperature to the kinetic temperature. Their approach was adopted by Hirata (2006). However, there are two main issues with their analysis that make an unambiguous application of their results questionable. The first is the form of the redistribution function for the resonance scattering in the Lyα line, and the second is the behavior of the solution of the radiative transfer equation.

The first point is that the actual redistribution function is always given as a linear combination of the coherent scattering in the atom's frame (leading to the $R_{II}$ redistribution function in the laboratory frame), and complete redistribution in the atoms' frame (see, e.g., Hubeny & Mihalas 2014, Chap. 10). The branching ratio is given by a ratio of the elastic (or almost elastic) collision rate to the spontaneous emission rate. DW85 implicitly assume that there is an equipartition of the upper state of Lyα according to their statistical weights, which means that there must be a sufficient rate of collisions between them, and therefore a non-negligible portion of complete redistribution.

The second point is that it must be recognized that the critical issue that influences the spin temperature is the color temperature in Lyα at large optical depths. It is well known from radiative transfer studies that even for a pure $R_{II}$ without a contribution of the complete redistribution part of the redistribution function, the source function is constant with frequency in the line core for line center optical depths in Lyα larger than about $10^3$, as first demonstrated by Hummer (1969); for discussion see Hubeny & Mihalas (2014, S15.6). If a portion of complete redistribution is present the constancy of the line source function extends to larger distances from the line center. This argues for the *constant* case.

We will revisit this in a future paper, but for now make all three options available in Cloudy, using the command "set Lya 21cm". Cloudy has, through C17, assumed the *excitation* case. Figure A1 compares physical conditions and the resulting spin temperature for these three cases for an environment similar to CNM of ISM with standard ISM metallicity and ISM radiation field. For this model, we consider hydrogen density = 100 cm$^{-3}$ and neutral hydrogen column density = $10^{20}$ cm$^{-2}$. The plots also show the gas kinetic temperature, typically ~30 K, and the Lyα excitation temperature, typically 3000 K, which are not affected by changes in the slope of the source function. The predicted 21 cm spin temperatures are more than 1 dex different. The spin temperature is close to the kinetic temperature in the "*kinetic*" case but is far higher than the kinetic temperature for the other two cases of the source function.

Tables:

Tabel 1. Einstein A coefficients and the occupation numbers for different transitions ($\Delta E$ is the energy difference between $^2P_{1/2}$ and $^2P_{3/2}$ levels in K).

| Transition | $A$ (s$^{-1}$) | Index |
|---|---|---|
| $_1S_{1/2}$ - $_0S_{1/2}$ | $2.85\times10^{-15}$ | 2→1 |
| $_1P_{1/2}$ – $_0S_{1/2}$ | $2.08\times10^{8}$ | 3→1 |
| $_1P_{1/2}$ – $_1S_{1/2}$ | $4.16\times10^{8}$ | 3→2 |
| $_1P_{3/2}$ – $_0S_{1/2}$ | $4.16\times10^{8}$ | 4→1 |

Table 2a. Kinetic temperature and 21 cm spin temperature for $N(\text{H I} = 10^{19})$ cm$^{-2}$

| $n_H$ (cm$^{-3}$) | 0.01 | 0.03 | 0.05 | 0.1 | 0.2 | 0.5 | 1 | 2 | 5 | 10 | 20 | 100 | 158 |
|---|---|---|---|---|---|---|---|---|---|---|---|---|---|
| Av. $T_s$ (K) | 2840 | 3090 | 3160 | 3250 | 3330 | 3350 | 3230 | 2870 | 1690 | 877 | 438 | 113 | 87 |
| $\tau_{21cm}$ | <10$^{-4}$ | <10$^{-4}$ | <10$^{-4}$ | <10$^{-4}$ | <10$^{-4}$ | <10$^{-4}$ | <10$^{-4}$ | <10$^{-3}$ | <10$^{-3}$ | 0.002 | 0.004 | 0.03 | 0.04 |
| Av. $T_k$ (K) | 9530 | 7990 | 7410 | 6470 | 5480 | 4030 | 3160 | 2290 | 997 | 468 | 244 | 88 | 73 |

Table 2b.  Kinetic temperature and 21 cm spin temperature for $N(\text{H I} = 10^{20})$ cm$^{-2}$

| $n_H$ (cm$^{-3}$) | 0.01 | 0.03 | 0.05 | 0.1 | 0.2 | 0.5 | 1 | 2 | 5 | 10 | 20 | 100 | 158 |
|---|---|---|---|---|---|---|---|---|---|---|---|---|---|
| Av. $T_s$ (K) | 2810 | 2990 | 3040 | 3100 | 3120 | 2930 | 2430 | 1600 | 573 | 222 | 115 | 53 | 47 |
| $\tau_{21cm}$ | <10$^{-3}$ | <10$^{-3}$ | <10$^{-3}$ | <10$^{-3}$ | 0.001 | 0.002 | 0.004 | 0.01 | 0.04 | 0.12 | 0.27 | 0.77 | 0.9 |
| Av. $T_k$ (K) | 9890 | 8480 | 7730 | 6320 | 4760 | 2010 | 757 | 340 | 149 | 93 | 64 | 36 | 33 |

Table 2c.  Kinetic temperature and 21 cm spin temperature for $N(\text{H I} = 10^{21})$ cm$^{-2}$

| $n_H$ (cm$^{-3}$) | 0.01 | 0.03 | 0.05 | 0.1 | 0.2 | 0.5 | 1 | 2 | 5 | 10 | 20 | 100 | 158 |
|---|---|---|---|---|---|---|---|---|---|---|---|---|---|
| Av. $T_s$ (K) | 2660 | 2820 | 2930 | 3090 | 3110 | 2220 | 1410 | 895 | 307 | 205 | 114 | 45 | 40 |
| $\tau_{21cm}$ | <10$^{-2}$ | <10$^{-2}$ | <10$^{-2}$ | <10$^{-2}$ | 0.01 | 0.03 | 0.09 | 0.2 | 0.9 | 1.6 | 3.2 | 11.5 | 14.7 |
| Av. $T_k$ (K) | 9930 | 8570 | 7820 | 6410 | 4750 | 1340 | 374 | 170 | 86 | 60 | 45 | 28 | 26 |

Table 2d. Kinetic temperature and 21 cm spin temperature for $N(\text{H I} = 10^{22})$ cm$^{-2}$

| $n_H$ (cm$^{-3}$) | 0.01 | 0.03 | 0.05 | 0.1 | 0.2 | 0.5 | 1 | 2 | 5 | 10 | 20 | 100 | 158 |
|---|---|---|---|---|---|---|---|---|---|---|---|---|---|
| Av. $T_s$ (K) | 2620 | 2800 | 2940 | 3100 | 2890 | 1760 | 1200 | 782 | 400 | 242 | 146 | 60 | 53 |
| $\tau_{21cm}$ | <0.1 | <0.1 | <0.1 | <0.1 | 0.14 | 0.6 | 1.3 | 2.54 | 6.3 | 11.9 | 21.7 | 59.2 | 67.7 |
| Av. $T_k$ (K) | 9940 | 8540 | 7740 | 6150 | 3570 | 501 | 229 | 139 | 86 | 66 | 55 | 46 | 45 |

FIGURES:

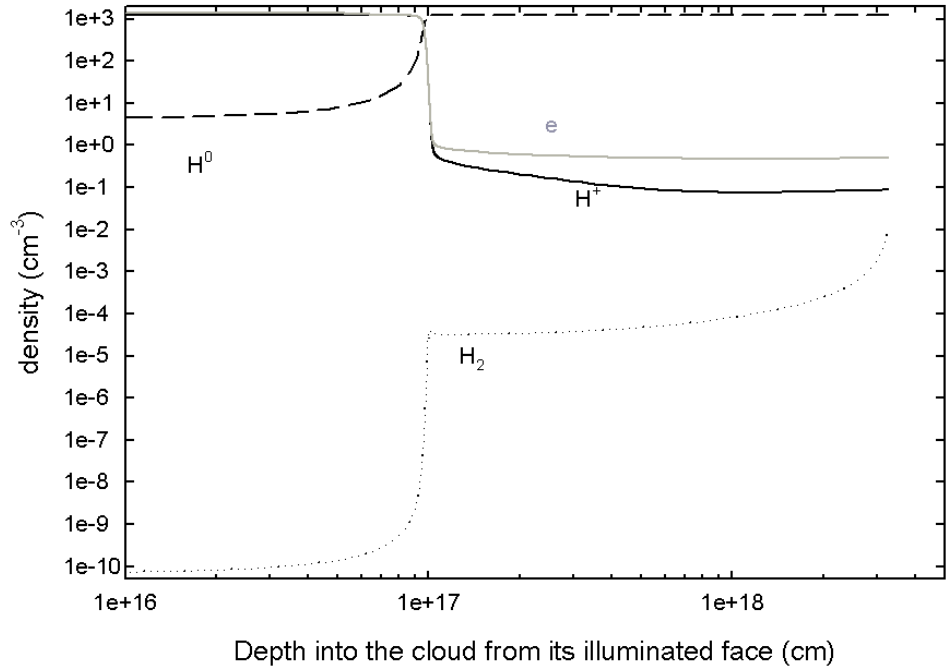

Figure 1. The solid (black), medium-dashed, dotted and solid (gray) lines represent the $H^+$, $H^0$, $H_2$, and $e^-$ densities as a function of depth into the cloud from its illuminated face for component B of Orion Veil. At the illuminated face, hydrogen is mostly $H^+$ (in the H II region); this is followed by an atomic hydrogen region.

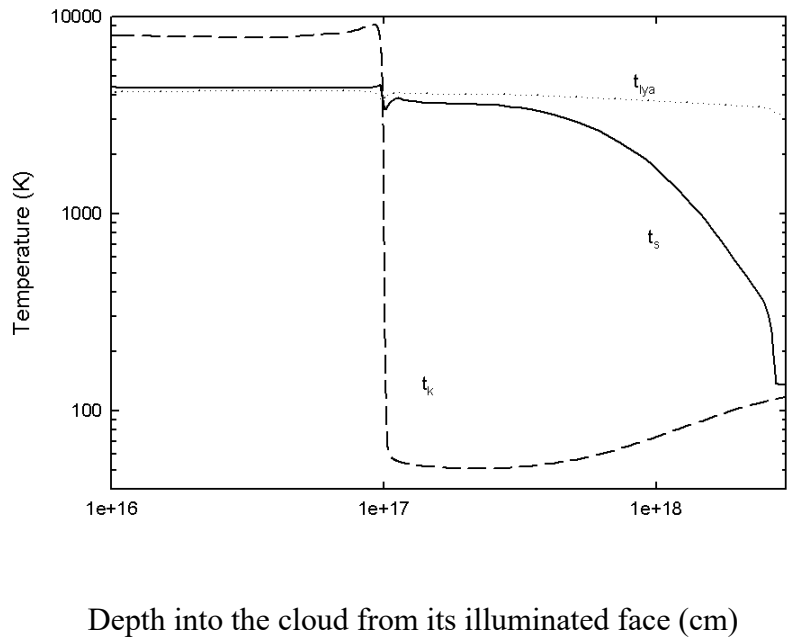

Depth into the cloud from its illuminated face (cm)

Figure 2. Various temperatures are plotted as a function of cloud depth for the component B of Orion Veil. The solid, dotted and medium-dashed lines represent $t_s$, $t_{lya}$ and $t_k$ respectively. This figure shows that $t_{lya}$ does not trace $t_k$.

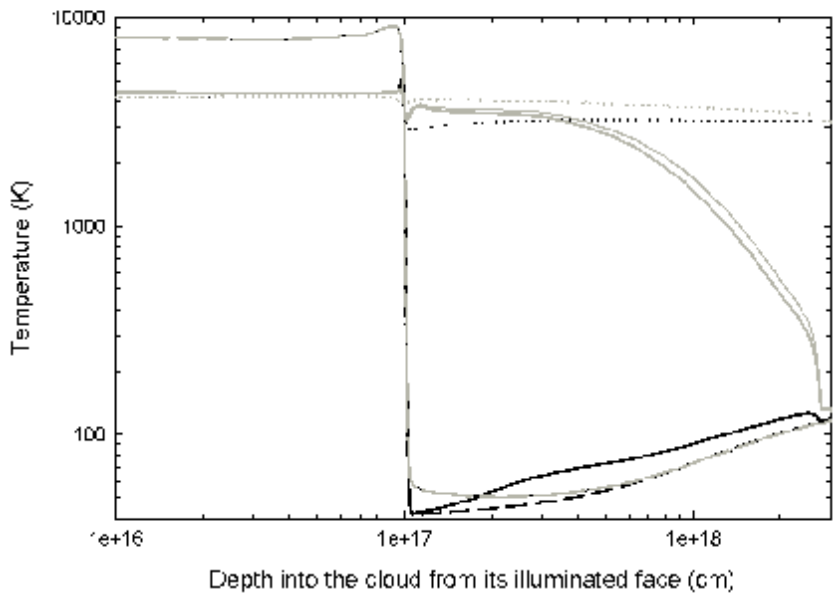

Figure 3. This figure shows the importance of continuum Lyα pumping on $t_s$. Various temperatures are plotted as a function of cloud depth for the Orion Veil. The solid, dotted and medium-dashed lines represent $t_s$, $t_{lya}$ and $t_k$ respectively as before. The corresponding thick-black, gray and thick-gray lines show effect of external Lyα continuum pumping with 0%, all in emission and 80% in emission respectively. Black and thick-gray lines overlap.

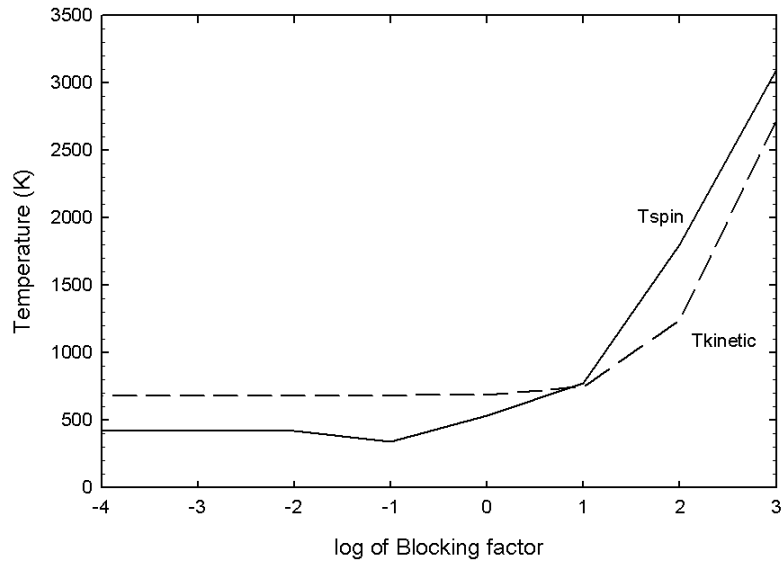

Figure 4. In real stars the Lyα line can be in absorption or in emission which has been denoted by Lyman alpha blocking factor in this figure. This figure shows the effect of Averaged Spin temperature (solid) and Kinetic temperature (medium-dashed) as a function of Lyman alpha blocking factor for Orion HII region like environment. Averaged Spin temperature is lower than the kinetic temperature for this model when this line is in absorption.

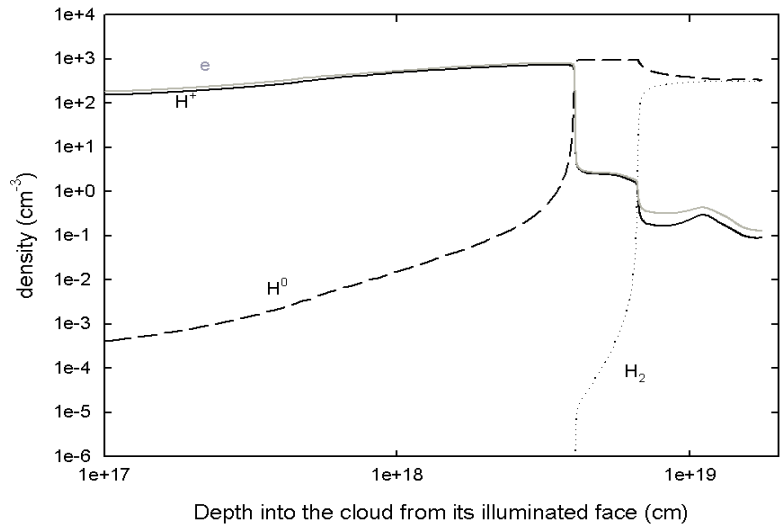

Figure 5. The solid (black), medium-dashed, dotted and solid (gray) lines represent the H$^+$, H$^0$, H$_2$, and e$^-$ densities as a function of depth into the cloud from its illuminated face for M17. Hydrogen is mostly H$^+$ in the H II region; this is followed by a neutral region (the PDR).

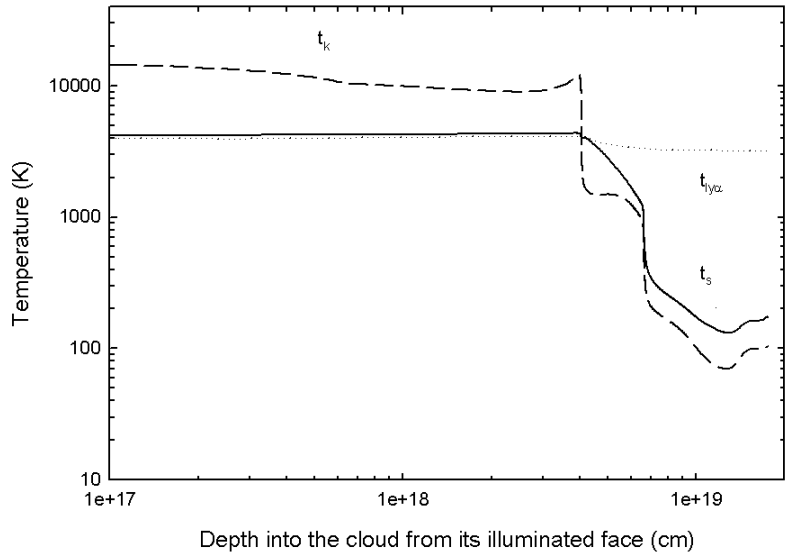

Figure 6. Various temperatures are plotted as a function of cloud depth for M17. The solid, dotted and medium-dashed lines represent $t_s$, $t_{lya}$ and $t_k$ respectively. This figure shows that $t_{lya}$ does not trace $t_k$.

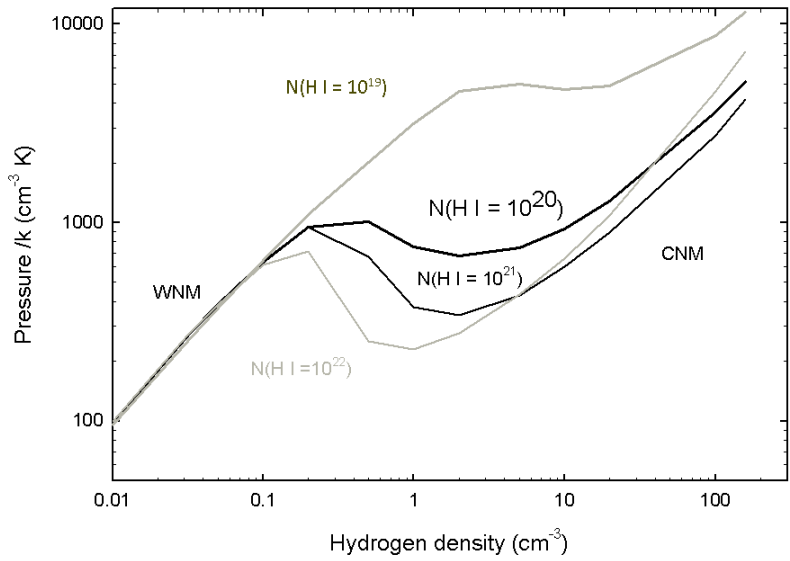

Figure 7. Above plot shows the difference between 21cm spin temperature (solid line) and the corresponding kinetic temperature (dashed-line) for various metallicities. Besides WNM and CNM, thermally unstable regions are also predicted (with a negative slop).

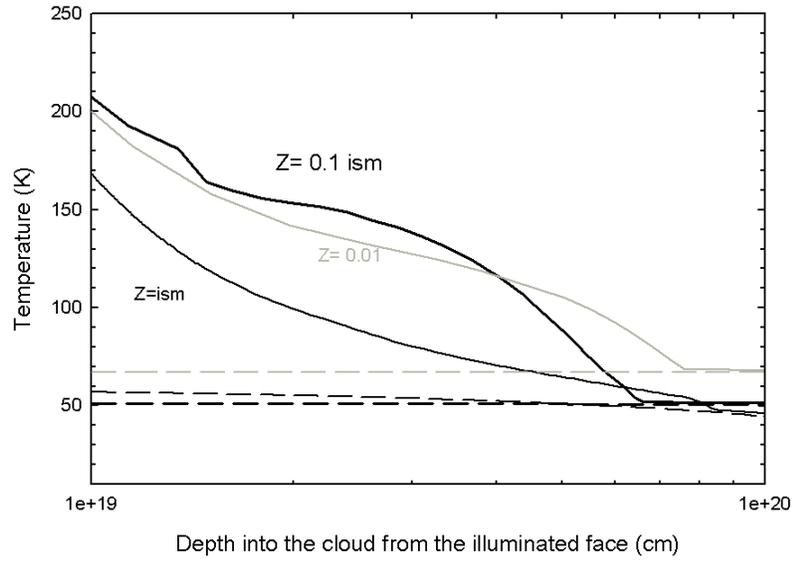

Figure 8a. Above plot shows the difference between $t_s$ (solid line) and the corresponding $t_K$ (dashed-line) for varying metallicity and dust content in the range of standard ISM to 0.01 ISM for a cloud with hydrogen density 10 cm$^{-3}$. Colors black, thick-black and gray represents ISM, 0.1 ISM and 0.01 ISM metallicity respectively.

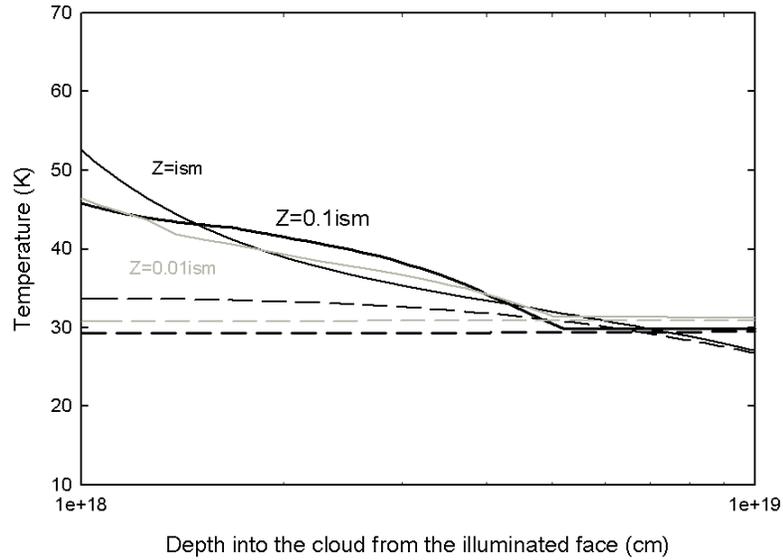

Figure 8b. Above plot shows the difference between $t_s$ (solid line) and the corresponding $t_K$ (dashed-line) for varying metallicity and dust content in the range of standard ISM to 0.01 ISM for a cloud with hydrogen density 100 cm$^{-3}$. Colors black, thick-black and gray represents ISM, 0.1 ISM and 0.01 ISM metallicity respectively.

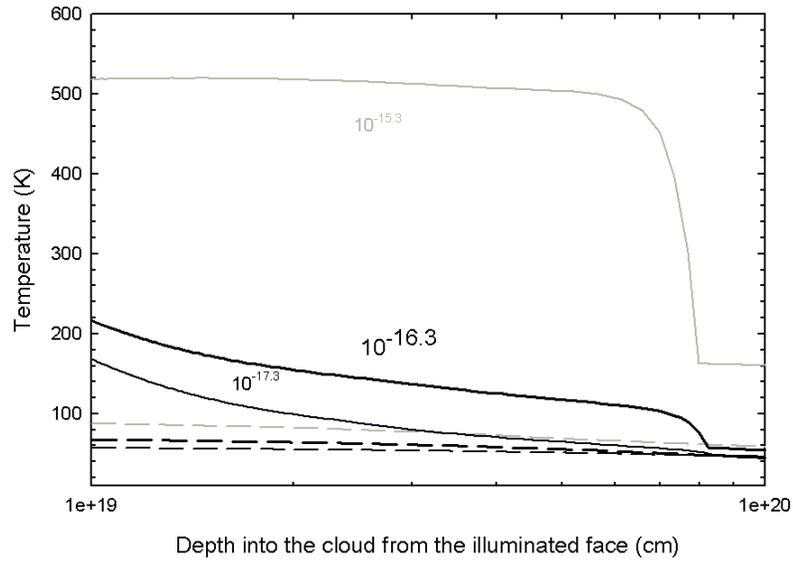

Figure 9a. Above plot shows the difference between 21cm spin temperature (solid line) and the corresponding kinetic temperature (dashed-line) for a cloud with hydrogen density 10 cm$^{-3}$ for various cosmic ray ionization rate of H$^0$ ranging from $10^{-17.3}$ s$^{-1}$ to $10^{-15.3}$ s$^{-1}$. Colors black, thick-black and gray represents $10^{-17.3}$, $10^{-16.3}$ s$^{-1}$ and $10^{-15.3}$ s$^{-1}$ CR rate respectively.

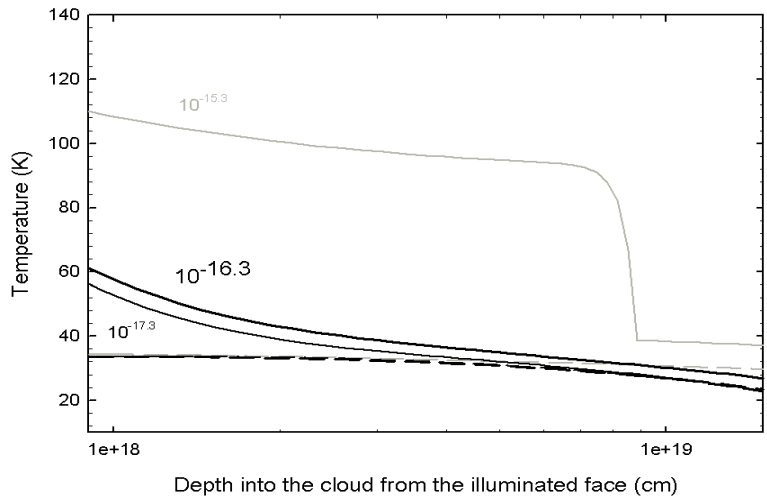

Figure 9b. Above plot shows the difference between 21cm spin temperature (solid line) and the corresponding kinetic temperature (dashed-line) for a cloud with hydrogen density 100 cm$^{-3}$ for various cosmic ray ionization rate of H$^0$ ranging from $10^{-17.3}$ s$^{-1}$ to $10^{-15.3}$ s$^{-1}$. Colors black, thick-black and gray represents $10^{-17.3}$, $10^{-16.3}$ s$^{-1}$ and $10^{-15.3}$ s$^{-1}$ CR rate respectively.

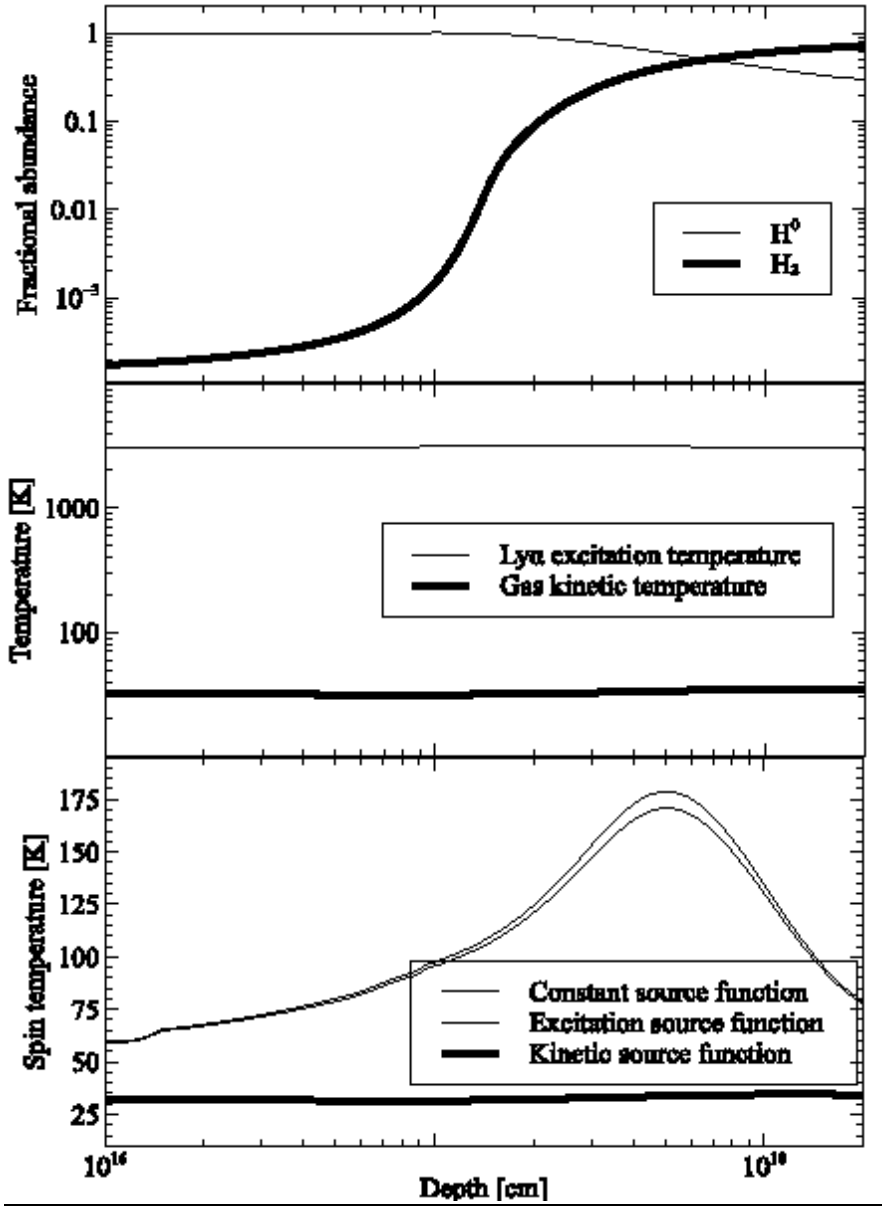

Figure 1A. Physical conditions and the resulting spin temperature for these three cases of Lyα Source function.